\documentstyle[12pt,aasms4,psfig]{article}
\newcommand{\E}[1]{$\times 10^{#1}$}
\def\msun{{\,M_\odot}}
\def\simlt{\lower.5ex\hbox{$\; \buildrel < \over \sim \;$}}
\def\simgt{\lower.5ex\hbox{$\; \buildrel > \over \sim \;$}}

\def\kms{{\rm\,km\;s^{-1}}}
\def\mdot{{\rm\,\msun\,yr^{-1}}}
\def\gms{{\rm\,g\,s^{-1}}}
\def\gcm3{{\rm\,g\,cm^{-3}}}
\def\ncm3{{\rm\,cm^{-3}}}

\def\K{{\rm\,K}}

\def\>{$>$}
\def\<{$<$}

\lefthead{Coker \& Melia}
\righthead{Spectrum of Sgr A*}

\begin{document}

\title{The Role of Magnetic Field Dissipation in the\\
       Black Hole Candidate Sgr A*}

\author{Robert F. Coker\altaffilmark{1}$^*$ and Fulvio Melia\altaffilmark{2}$^{*\dag}$}
\affil{$^*$Physics Department, The University of Arizona, Tucson, AZ 85721}
\affil{$^{\dag}$Steward Observatory, The University of Arizona, Tucson, AZ 85721}

\altaffiltext{1}{NASA GSRP Fellow.}
\altaffiltext{2}{Sir Thomas Lyle Fellow.}

\begin{abstract}
The compact, nonthermal radio source Sgr A* at the Galactic Center appears
to be coincident with a $\sim 2.6\times 10^6\;M_\odot$ point-like object.
Its energy source may be the release of gravitational energy as gas
from the interstellar medium descends into its deep potential well.
However, simple attempts at calculating the radiative spectrum and flux
based on this picture have come tantalizingly close to the observations,
yet have had difficulty in accounting for the unusually low efficiency
in this source.  Regardless of whether the radiating particles in the
accretion flow are thermal or nonthermal, there now appear to be two
principal reasons for this low conversion rate of dissipated energy into
radiation: (1) the plasma separates into two temperatures, with the protons
attaining a significantly higher temperature than that of the radiating electrons,
and (2) the magnetic field {\bf B} is sub-equipartition, which reduces the magnetic
bremsstrahlung emissivity, and therefore the overall power of Sgr A*.  In 
this paper, we investigate the latter with a considerable improvement over
what has been attempted before. In particular, rather than calculating
{\bf B} based on some presumed model (e.g., equipartition with the thermal
energy of the gas), we instead infer its distribution with radius empirically with
the requirement that the resulting spectrum matches the observations.  Our
assumed ansatz for {\bf B}(r) is motivated in part by earlier calculations of the
expected magnetic dissipation rate due to reconnection in a compressed flow. 
We find reasonable agreement with the observed spectrum of Sgr A* as long
as its distribution consists of 3 primary components: an outer equipartition
field, a roughly constant field at intermediate radii ($\sim 10^3$ Schwarzschild
radii), and an inner dynamo (more or less within the last stable orbit for
a non-rotating black hole) which increases {\bf B} to about 100 Gauss.  The
latter component accounts very well for the observed sub-millimiter hump in this source.
\end{abstract}


\section{Introduction}
The Galactic center (GC) has long been suspected of harboring a central
mass concentration, which appears to be coincident with the unique,
nonthermal radio source, Sgr A*. Haller et al. (1996) used the velocity
dispersions of stars at $\ga 0.1$ pc from Sgr A* to derive a compact mass
of $\sim 2\times 10^6 \;M_\odot$. This is consistent with the value
of $\sim 2.5-3.2\times 10^6\;M_\odot$ derived more recently by Genzel et al.
(1996), using the radial velocities and velocity dispersions of
$\sim 25$ early-type stars and of $\sim 200$ red giants and
supergiants within the central 2 pc.  A third technique for tracing the
central gravitational potential
is based on the acquisition of proper motions for the $\sim
50-100$ brightest stars within the radial range $\sim 0.004-0.4$ pc
(Eckart \& Genzel 1996; Ghez et al. 1998).  These stellar motions seem to 
require a central dark mass of $(2.6\pm 0.2)\times 10^6\;M_\odot$, 
in good agreement with earlier ionized gas
kinematics and the velocity dispersion measurements.

Of course, showing that the GC must contain a centralized mass
concentration does not necessarily imply that this dark matter
is in the form of a compact object with a few million solar masses.
It does not even imply that the unusual radio source Sgr A* must
be associated with it.  VLBA images of Sgr A* with milliarcsecond
resolution (Bower \& Backer 1998) show that at $\lambda$7 mm, its size is
$0.76\pm0.04$ mas, or roughly $6.2\times 10^{13}$ cm, much smaller
than $\sim 0.01$ pc, the present limiting region within which the $2-3\times
10^6\;M_\odot$ are contained.  So the dark matter may be distributed,
perhaps in the form of white dwarfs, neutron stars, or $\sim
10\;M_\odot$ black holes (e.g., Haller et al. 1996), though
the latest stellar kinematic results appear to rule out the first two
possible constituents (Genzel et al. 1996).

Whatever the composition of a distributed mass concentration is,
one is left with the task of accounting for the nature of
Sgr A* itself.  It is likely that many of Sgr A*'s characteristics
are associated with the liberation of gravitational energy as gas
from the ambient medium falls into a central potential well
(Melia 1994;  see also Ozernoy 1989 for an alternative conclusion
regarding wind accretion).  There is ample observational evidence in this region
for the existence of rather strong winds in and around Sgr A* itself
(from which the latter is accreting),
e.g., the cluster of mass-losing, blue, luminous stars comprising the
IRS 16 assemblage located within several arcseconds from the nucleus.
Measurements of high outflow velocities associated with IR sources in
Sgr A West (Krabbe et al. 1991) and in IRS 16 (Geballe et al. 1991),
the $H_2$ emission
in the circumnuclear disk (CND) from molecular gas being shocked by a
nuclear mass outflow (Genzel et al. 1996; but see Jackson et al. 1993 for the
potential importance of UV photodissociation in promoting this $H_2$
emission), broad Br$\alpha$, Br$\gamma$ and He I emission lines from
the vicinity of IRS 16 (Hall et al. 1982;
Allen et al. 1990; Geballe et al. 1991), and
radio continuum observations of IRS 7 (Yusef-Zadeh \& Melia 1991),
provide clear
evidence of a hypersonic wind, with a velocity $v_w \sim500-1000$ km
s$^{-1}$, a number density $n_w\sim10^{3-4}$ cm$^{-3}$, and a total
mass loss rate $\dot M_w\sim\times10^{(-3)-(-4)}\;\dot M_\odot$, pervading
the inner parsec of the Galaxy.
Even so, the observations do not yet provide sufficient information
for us to identify the physics of accretion when the infalling gas
penetrates to within about $10^3$ or $10^4$ Schwarzschild radii of
the central object. 

\subsection{Behavior of the Accreting Gas at Small Radii}
Three-dimensional hydrodynamic simulations (Coker \& Melia 1997) indicate that
the accreted specific angular momentum $\lambda$ (in units of $cr_s$,
where $r_s\equiv 2GM/c^2$ is the Schwarzschild radius in terms of
the black hole mass $M$) can vary by 
$50\%$ over $\simlt$ 200 years with an average equilibrium value for $\lambda$ of 
$40 \pm 10$. Thus, even with a possibly large amount of angular momentum 
present in the wind, relatively little specific angular momentum is 
accreted.  This is understandable since clumps of gas with a high specific 
angular momentum do not penetrate to within 1 $R_A$, where 
\begin{equation}\label{eq:radef}
R_A\equiv 2GM/{v_w}^2
\end{equation}
is the capture radius defined in terms of the wind
velocity $v_w$ at infinity.  The variability in the sign of the 
components of $\lambda$ suggests
that if an accretion disk forms at all, it dissolves, and reforms (perhaps)
with an opposite sense of spin on a time scale of $\sim 100$ years.

The captured gas is highly ionized and magnetized, so it radiates via
brems\-strahlung, cyclo-synchrotron and inverse Compton processes.
However, for purely spherical accretion, the efficiency of 
converting gravitational energy into
radiation is quite small (as little as $10^{-4}$ in some cases), so most
of the dissipated energy is carried inwards (Shapiro 1973;  Ipser \& Price 1977;
Melia 1992).  In fact, if the magnetic field is a negligible fraction
of its equipartition value (see below), Sgr A* would be undetectable
at any frequency, except perhaps at soft X-ray energies. But as the plasma
continues to compress and fall toward smaller radii,
one or more additional things can happen, each of which corresponds
to a different theoretical assumption, and therefore a potentially
different interpretation.

The questions one may ask include the following: (1) Does the flow
carry a large specific angular momentum so that it forms a disk
with lots of additional dissipation?  (2) Does the flow produce a radiatively
dominant non-thermal particle distribution at small radii (e.g.,
from shock acceleration), or does thermal emission continue to
dominate the spectrum?  (3) Does the flow lead to an expulsion of plasma
at small radii that forms a non-thermal jet, which itself may then
dominate the spectrum?  These, either individually or in combination,
have led to a variance of assumptions about the nature of the inflowing
gas that then form the basis for the development of different interpretations.

Observationally, a key issue is why the infalling gas maintains a low
radiative efficiency.  Beckert \& Duschl (1997) suggest that shocks
in the accreting plasma produce a power-law electron distribution,
which is truncated by strong cooling.  This forms a ``quasi'' mono-energetic
distribution.  The overall emission, which is strictly non-thermal,
is suppressed by constraining the number density of relativistic
particles and the intensity of the magnetic field (at about 5 to 10
Gauss).

Falcke, Mannheim and Biermann (1993) and Falcke \& Biermann (1999), 
on the other hand, assume that
the infalling plasma eventually produces a jet of power-law electrons
whose number density varies with radius in the expulsion.  The overall
emission, which is a sum of non-thermal components, is also
suppressed by constraining the particle number density and hence
the equipartition magnetic field, both of which are assumed to be
scaled by a low luminosity disk.

In the picture developed by Narayan, et al. (1998), the infalling
gas is assumed to carry a very large angular momentum, so that a disk
forms with an outer radius at more than $10^5$ Schwarzschild radii.  To
suppress the overall emission, which now includes the additional
dissipation of this large angular momentum, it is also assumed that
the electron temperature is much lower than that of the protons ($T_e\ll
T_p$).  In fact, $T_e<10^{10}$ K.  Since the electrons do the radiating,
the efficiency remains small even though the protons are very hot.

\subsection{A Sub-equipartition Magnetic Field}
The idea that Sgr A*'s low radiating efficiency is due to a sub-equipartition
magnetic field $B$ deserves further attention, especially in view of the 
fact that the actual value of $B$ depends strongly on the mechanism of field 
line annihilation, which is poorly understood.  Two processes that have been
proposed are (i) the Petschek (1964) mechanism, in which dissipation of the
sheared magnetic field occurs in the form of shock waves surrounding special
neutral points in the current sheets and thus, nearly all the dissipated 
magnetic energy is converted into the magnetic energy carried by the emergent 
shocks; and (ii) van Hoven's (1979) tearing mode instability, which relies 
on resistive diffusion of the magnetic field and is very sensitive to the 
physical state of the gas.  In either case, the magnetic field dissipation rate
is a strong function of the gas temperature and density, so that assuming
a fixed ratio of the magnetic field to its equipartition value may not
be appropriate.

Kowalenko \& Melia (1999) have used the van Hoven prescription to calculate
the magnetic field annihilation rate in a cube of ionized gas being compressed
at a rate commensurate with that expected for free-fall velocity onto the
nucleus at the Galactic Center.  
Whereas the rate of increase $\partial B/\partial t|_f$ in
$B$ due to flux conservation depends only on the rate $\dot r$ of the gas,
the dissipation rate $\partial B/\partial t|_d$ is a function
of the state variables and it is therefore not necessarily correlated with
$\dot r$.  Although these attempts at developing a physical model for
magnetic field dissipation in converging flows is still rather simplistic,
it is apparent from the test simulations that the equipartition assumption is not
always a good approximation to the actual state of a magnetohydrodynamic flow,
and very importantly, that the violation of equipartition can vary in degree
from large to small radii, in either direction.  As such, calculations that assume 
equipartition of the magnetic field with the gas throughout the domain of 
solution may be greatly underestimating the importance of the deviations of $B$ 
from its $B_{eq}$ value since the predicted spectrum relies critically on the 
contribution from magnetic bremsstrahlung.

\subsection{Impact on Sgr A*'s Spectrum}
The first serious attempt at modeling the spectrum of Sgr A* as being due to
emission by the accreting gas was carried out by Melia (1992, 1994), who
assumed a black hole mass of 
$\approx 10^6\;M_\odot$.  But this mass is no longer consistent with the
now more accurately known value of $\sim 2.6\times 10^6\;M_\odot$, which
accounts for a factor of $\sim 7$ increase in $\dot M$.  In addition, the earlier
calculations integrated the cyclo-synchrotron emissivity out to the lowest
20 harmonics only, which misses some of the contribution to the flux by
the highest temperature gas at the smallest radii (Mahadevan, Narayan \& Yi
1996).  In this paper, we recalculate the spectrum
produced by a quasi-spherical infall onto Sgr A* using the updated mass
value, a more accurate handling of the magnetic bremsstrahlung and an empirical
fit to the magnetic field, motivated by the simulations of magnetic dissipation
discussed above.  In section \S\ 2, we derive the equations
governing this spherical infall, which we adopt as a simplified version
of the real accretion picture.  Of course, the real
accretion flow will deviate from radial at small distances from the black
hole, where the gas begins to circularize with its advected specific angular
momentum.  In a fully self-consistent calculation, we will use the
distributions derived from an actual 3D hydrodynamic simulation as the
basis for calculating the emissivity.  The model parameters and
results of our calculations are discussed in \S\ 3, and we summarize
our conclusions in \S\ 4.

\section{Equations Governing Spherical Accretion}\label{sec:speceq}
\subsection{The Radial Profiles}
We follow the sequence of derivations in Shapiro (1973), with the primary 
differences being the inclusion of the magnetic field and a radiation
pressure term and the fact that we restrict out attention to
supersonic flows.  The equation of mass conservation reduces to
\begin{equation}\label{eq:mdoteq}
\dot M = 4\pi r^2\rho v\;,
\end{equation}
where $\dot M$ is the mass accretion rate onto the black hole, $\rho$ is the mass
density of the accreting gas, $-v\equiv u^r$ is the radial component of the fluid 4-velocity
(but defined to be positive inwards, so that $v=-{\rm d}r/{\rm d}\tau$ for
the infalling plasma), and $r$ is the distance from the black hole.
Equation (\ref{eq:mdoteq}) can be recast into the form
\begin{equation}\label{eq:nprime}
{{n^\prime }\over{n}} = -\left({{v^\prime}\over{v}} + {{2}\over{r}}\right)\;,
\end{equation}
where a prime denotes ${\rm d}/{\rm d}r$.  At $R_0$, where the numerical integration 
begins, we assume that the gas is non-relativistic and that the gravitational 
potential is weak.

The second equation, arising from momentum conservation, is the steady state 
relativistic Euler equation for a 
spherical geometry:
\begin{equation}\label{eq:Euler}
vv^\prime = -\left({{c^2 + v^2 - 2GM/r}\over{P +
e_{\rho} + \epsilon }}\right){P^\prime}_{th} - {{GM}\over{r^2}}\;,
\end{equation}
where the mass of the central black hole is given by $M$, the total pressure is given by
\begin{equation}
P = {{B^2}\over{8\pi}} + P_{th}\;,
\end{equation}
the non-magnetic pressure is given by
\begin{equation}
P_{th} = 2nkT + P_{rad}\;,
\end{equation}
the particle mass-energy density is 
\begin{equation}
e_{\rho} = m_p c^2 n\;,
\end{equation}
and the internal energy density of the gas is 
\begin{equation}
\epsilon = \alpha nkT + 3 P_{rad} + {{B^2}\over{8\pi}}\;.
\end{equation}
In the fully ionized but non-relativistic limit (i.e., $10^5 < T < 6\times10^9$ K),
$\alpha=3$.  On the other hand, in the relativistic electron limit 
($6\times10^9< T < 10^{13}$ K), $\alpha=9/2$.  We use the general expression 
from Chandrasekhar (1939) that is valid for all $T$: 
\begin{equation}
\alpha = 3 + x \left( {{3\K_3(x) + \K_1(x)}\over{4 \K_2(x)}} - 1 \right)
+ y \left( {{3\K_3(y) + \K_1(y)}\over{4 \K_2(y)}} - 1 \right) \;,
\end{equation}
where $x \equiv m_e c^2 / kT$, $y \equiv m_p c^2 / kT$ and K$_i$ refers to the
i$^{th}$ order modified Bessel function.
We assume that the gas consists solely of completely ionized hydrogen.
Note that since we will assume a mostly radial $B$ (see below), the numerator of Equation
(\ref{eq:Euler}) does not depend on the magnetic field.  That is, there is no large
scale current, since $\bf{v}\times\bf{B} = 0$.  The radiation pressure, $P_{rad}$, 
is given by the Rayleigh-Jeans approximation
\begin{equation}
P_{rad} = {{8\pi}\over{9}} kT {\left({{\nu_m}\over{c}}\right)}^3\;,
\end{equation}
where $\nu_m$ is the frequency below which the radiative emission is 
highly absorbed, so that the optical depth, $\tau_r^\infty(\nu_m)$, 
from $r$ to infinity is unity.  This assumes that the accretion flow is 
relatively unimpeded by radiation pressure, or,
in other words, that the accretion is sub-Eddington.  As mentioned in Melia (1992),
with $L_{Sgr A*} \sim 10^5 L_\odot \ll L_{Ed} \sim 10^{11} L_\odot$,
(Zylka et al. 1995; Shapiro and Teukolsky 1983),
this certainly appears to be the case for Sgr A*.
Numerically, this means that $h\nu_m \ll 
kT$, which we verify {\sl a posteriori}.  

The third primary equation follows from energy conservation, which we derive from the
first law of thermodynamics:
\begin{equation}\label{eq:1st}
{{\rm d}\over{{\rm d}\tau}}\left({{e_{\rho} + \epsilon}\over{n}}\right) =
-P_{th} {{\rm d}\over{{\rm d}\tau}}\left({{1}\over{n}}\right) + {{\Gamma - \Lambda}\over{n}}\;,
\end{equation}
where $\tau$ is the proper time in the gas frame.
The use of $P_{th}$ rather than $P$ assumes that the compression of the gas is
parallel to the magnetic field lines (i.e., radial inflow with a mostly radial $B$).
The heating ($\Gamma$) and cooling ($\Lambda$) terms render the flow non-adiabatic 
(i.e., ${\rm d}s/{\rm d}\tau \neq 0$).  The radiative cooling includes
magnetic bremsstrahlung and electron-ion and electron-electron thermal bremsstrahlung 
(see Melia 1994, with an updated prescription in Melia \& Coker 1999).  Local UV heating 
from nearby stars results in a minimum gas temperature of $10^{4-5}$K (Tamblyn et al. 1996),
but the shocked gas at the model's outer radius, $R_0$, is expected to be
hotter.  Thus, the major heating term (other than the effects of
compression) is expected to be due to magnetic field reconnection.  
Specifically, we use (see Ipser \& Price 1982)
\begin{equation}
\Gamma = {{n v}\over{8 \pi}} \left\{\left({{B^2}\over{n}}\right)^\prime - {{B^2}\over{n}} 
\left({{v^\prime}\over{v}}-{{2}\over{r}}\right) \right\}\;.
\end{equation}
Thus, if the magnetic field is flux conserved, for which $B(r) \propto r^{-2}$, then
no reconnection is taking place and $\Gamma=0$.  However, if the magnetic field remains
in approximate kinetic equipartition with the gas, meaning that its pressure increases in tandem 
with the ram pressure of the accreting gas ($P_{mag}/P_{ram} \sim$ constant), then $B(r) 
\propto r^{-5/4}$, and $\Gamma\not=0$.  Since $T(r)$, $v(r)$, and $n(r)$ will
turn out not to be perfect power-laws, these statements are only approximate.

To find the velocity gradient, $v^\prime$, we place the above definitions for 
$P_{th}$ and $P_{rad}$ into Equation (\ref{eq:Euler}) to get
\begin{equation}
vv^\prime = -\left({{c^2 + v^2 - 2GM/r}\over{P + e_{\rho} + \epsilon }}\right)\left(2k[n^\prime 
T + nT^\prime] + {{8k\pi}\over{9c^3}}[T^\prime \nu^3_m + T 3\nu^2_m {\nu^\prime}_m]\right) - 
{{GM}\over{r^2}}\;.
\end{equation}
Thus, substituting for $n^\prime$ from Equation (\ref{eq:nprime}), we have,
\begin{equation}\label{eq:vprime}
v^\prime = {{-H v \left[2nk(T^\prime - {{2T}/{r}}) +{{(8k\pi}/{9c^3}})(T^\prime \nu^3_m 
+ T 3\nu^2_m {\nu^\prime}_m)\right] - v{{GM}/{r^2}}} \over{v^2-2nkTH}}\;,
\end{equation}
where, for ease of writing, we have defined the quantity 
\begin{equation}
H\equiv{{c^2 + v^2 - 2GM/r}\over{P + e_{\rho} + \epsilon }}\;.
\end{equation}
Note that in the non-relativistic and small $B$ limit, $H = 1/\rho$.
For simplicity, we write $v^\prime$ as the sum of two terms:
\begin{equation}
v^\prime = f T^\prime + g\;,
\end{equation}
where
\begin{equation}
f \equiv {{-H v k\left(2n + {{8\pi\nu^3_m}/{9c^3}}\right)}\over{v^2-2nkTH}}\;,
\end{equation}
and
\begin{equation}
g \equiv {{-H v kT\left({{-4n}/{r}} +{{8\pi\nu^2_m{\nu^\prime}_m}/{3c^3}}\right) - 
v{{GM}/{r^2}}} \over{v^2-2nkTH}}\;.
\end{equation}

The form of these expressions is that of the classic wind equations (see, Parker 1960; Melia 1988).
In the simulations we consider here, the gas is supersonic at $R_0$ and remains
supersonic on its inward trajectory (i.e., for $r< R_0$).  We, therefore, avoid
the special handling required for solutions that cross any sonic points, where 
the denominator of Equation (\ref{eq:vprime}) vanishes.

Since ${\rm d}(e_\rho/n)/{\rm d}\tau = 0$, Equation (\ref{eq:1st}) reduces to
\begin{equation}
{{\rm d}\over{{\rm d}\tau}}\left({{\epsilon}\over{n}}\right) =
-P_{th} {{\rm d}\over{{\rm d}\tau}}\left({{1}\over{n}}\right) + {{\Gamma - \Lambda}\over{n}}\;.
\end{equation}
Substituting in $P_{th}$ and $\epsilon$, we get
\begin{equation}
-v{{\rm d}\over {\rm d}r}\left(\alpha kT + {{8\pi\nu^3_m kT}\over{3nc^3}} + {{B^2}\over{8\pi n}}\right) =
-v\left(-2nkT - {{8\pi\nu^3_m kT}\over{9c^3}} \right){{-n^\prime}\over{n^2}}
+ {{\Gamma-\Lambda}\over{n}}\;.
\end{equation}
Substituting in Equation (\ref{eq:nprime}) and explicitly taking the derivatives, 
we then obtain
\begin{eqnarray}\label{eq:temp}
T^\prime\left(\alpha k + {{8\pi\nu^3_m k}\over{3nc^3}}\right) = -\left({{v^\prime}\over{v}}
+ {{2}\over{r}}\right)
\left(2kT + {{32\pi\nu^3_m kT}\over{9nc^3}} + {{B^2}\over{8\pi n}}\right) \nonumber \\
+ {\Lambda-\Gamma\over{nv}} - {{BB^\prime}\over{4\pi n}}
- {{8\pi\nu^2_m kT{\nu_m}^\prime}\over{nc^3}} \;.
\end{eqnarray}

We use Equations (\ref{eq:vprime}) and (\ref{eq:temp}) to determine $v^\prime$
and $T^\prime$, respectively,
then solve for $v$ and $T$ using an
implicit differencing scheme, with
\begin{equation}
T_{j+1} = T_{j} + (r_{j+1}-r_{j}) T^\prime_{j+1}\;,
\end{equation}
where $T^\prime_{j+1}$ is a function of $T_j, v_j, r_j, B_j,$ and $v^\prime_j$.  
Similar relations hold for $v$.  Thus, the
whole flow is determined given some outer boundary conditions ($R_0$, $\beta$, $f_v$, and
$f_T$; see below), the empirically inferred behaviour of $B(r)$, and a prescription
for finding $\nu_m$.  Note that boundary condition values for $T^\prime$, $v^\prime$,
$B^\prime$, and $\nu_m^\prime$ are also needed; at the outer boundary, we assume the flow
is optically thin, in adiabatic free-fall with a thermal equipartition field and determine
the derivatives accordingly.

\subsection{The Optical Depth and a Prescription for $\nu_m$}
Since the gas is not expected to be optically thin at
low radio frequencies, the effective optical depth
as a function of frequency must be determined before the radial profiles can be calculated.
For Sgr A*, we assume (and check {\sl a posteriori})
that the energy absorbed at any given radius is small compared to
the thermal and kinetic energy at that radius; this is not likely to be true for 
objects that are accreting closer to their Eddington luminosity.  Further, we
assume that multiple scatterings are unimportant.

Following Rybicki \& Lightman (1979), we use, for an effective optical depth in
a zone of observed size ${\rm d}r_0$ ($=r_{j+1} - r_j$) at infinity,
\begin{equation}\label{eq:taueq}
\tau_j = {\rm C d}r
\sqrt{\alpha_{abs}(\alpha_{abs} + n \sigma_{scat})},
\end{equation}
where
\begin{equation}\label{eq:rprime}
{\rm d}r = {\rm d}r_0 {{1-\beta \bar \mu}\over{\sqrt{(1-{{2GM}/{rc^2}})(1-\beta^2)}}}\;,
\end{equation}
$\alpha_{abs}$ is the absorption coeffecient,
$n$ is the electron number density, and $\sigma_{scat}$ is the
electron scattering cross section.  Note that $r_{j=1} = 1 r_s = 2 GM/c^2$.
The coefficient C in Equation (\ref{eq:taueq}) is a geometric term arising from the fact that in 
spherical symmetry, the average path length of a photon that reaches the 
observer is somewhat larger than ${\rm d}r_0$.  It is given by
\begin{equation}
{\rm C} = \min\{ {{1+\sqrt{3}}\over{2}} , {{2}\over{1-\mu_{max}}} \} \;,
\end{equation}
where $\mu_{max}$ is defined in Equation (\ref{eq:mumax}).
In Equation (\ref{eq:rprime}),
\begin{equation}
\beta = {{v}\over{c \sqrt{1 + (v/c)^2 - {{2GM}/{rc^2}} } } }
\end{equation}
and $\bar \mu$ is the average of the minimum of the cosine of the
angle between the line of sight and the flow (which is here $-1$) and the maximum,
\begin{equation}\label{eq:mumax}
\left|\mu^{max}\right| = \sqrt{ {{27}\over{4}} \left({{2GM}\over{rc^2}}\right)^2
\left({{2GM}\over{rc^2}} - 1\right) + 1 } \;\;,
\end{equation}
given by Zel\'dovich \& Novikov (1971).
This maximum in $\mu$ is a consequence of the fact that not all of the emitted photons
reach the observer; some are captured by the black hole.  Note that $\mu_{max}$
changes sign and becomes negative at radii smaller than 1.5 $r_s$.  Also, $\mu_{max} 
\rightarrow -1$ at the event horizon since the curvature in the trajectory
of photons emitted at other angles takes them back into the event horizon.
Although $\tau$ has a more complex angular dependence, the use of $\bar \mu$ in 
Equation (\ref{eq:rprime}) is a necessary simplification at this point.

For $\alpha_{abs}$, we use Kirchoff's Law,
\begin{equation}
\alpha_{abs} = j_{\nu} / B_{\nu}\;,
\end{equation}
where $j_{\nu}$ is the total emissivity (in ergs cm$^{-3}$ s$^{-1}$ Hz$^{-1}$
steradian$^{-1}$) and
$B_{\nu}$ is the blackbody Planck function.  For the models discussed here,
we use a total emissivity that includes cyclo-synchrotron emission 
(Coker \& Melia 1999) and electron-ion and electron-electron bremsstrahlung 
(Melia \& Coker 1999).  For $\sigma_{scat}$ we use
the exact scattering cross section (see, e.g., Lang 1980)
\begin{equation}
\sigma_{scat} = {{3}\over{4}} \sigma_T \left[ {{1+x}\over{x^3}}
\left( {{2x(1+x)}\over{1+2x}}-ln(1+2x) \right) + {{1}\over{2x}} ln(1+2x)
- {{1+3x}\over{(1+2x)^2}} \right]\;,
\end{equation}
where $x = \gamma h \nu / (m c^2)$ and 
\begin{equation}
\gamma = \max\{ 1, \sqrt{12} kT / mc^2\}  
\end{equation}
is the RMS value of the thermal velocity at the temperature in the given zone.

Finally, to find the total optical depth from zone $j$ out to infinity at
some observed frequency $\nu_0$, related to the emitted frequency $\nu$ by
\begin{equation}
\nu_0 = \nu\, {{\sqrt{(1-{{2GM}/{rc^2}})(1-\beta^2)}}\over{1-\beta \bar \mu}}\;,
\end{equation}
we use 
\begin{equation}
\tau_{\nu_0}^\infty = \sum_{k=j+1}^{k=\infty} (r_j/r_k)^2 \tau_k \;. 
\end{equation}
The minimum frequency $\nu_m$ that a photon needs to have in order to escape 
is found by iteratively determining the wavelength at which $\tau_k$ is unity 
with the caveat that $\nu_m$ not be less than the plasma frequency 
\begin{equation}
\nu_p = e \sqrt{ {{n}\over{\pi m_e}} }\;.
\end{equation}
This caveat is required since photons with a frequency less than $\nu_p$
are unable to propogate and are thus trapped by the infalling gas.
Once the radial profiles (optical depth, density, velocity, and temperature)
are
determined, it is possible to calculate the emission spectrum for
a given magnetic field profile.

\subsection{Calculation of the Spectrum}
Now we are ready to calculate the predicted observable luminosity
$L_{\nu_0}$ at infinity (see Shapiro 1973; Ipser \& Price 1982; and applied to
Sgr A* in Melia 1992, 1994):
\begin{equation}
L_{\nu_0} = 8 \pi^2 \sum_{j=1}^{j=J} e^{-\tau_{\nu_0}^\infty(j)} r_j^2 
\left( 1 - \beta^2 \right) \int_{-1}^{\mu_{max}}
{{d\mu}\over{(1-\beta \mu)^2}} I_{\nu}\;,
\end{equation}
where, if the emitting zone is optically thick (i.e., if $\tau_j > 1$), then
\begin{equation}
I_{\nu} = B_{\nu} \left( 1 - e^{-\tau_j} \right)\;.
\end{equation}
Otherwise
\begin{equation}
I_{\nu} = j_{\nu} e^{-\tau_j} {\rm d}r\;.
\end{equation}
In these two expressions,  $B_\nu$ and $j_\nu$ depend on $\mu$ not $\bar \mu$.  The use
of $\tau_j$ here is a calculational simplification; ideally, it too should be a function
of $\mu$.  The sum over $j$ is truncated at $J$, for which $r_J \equiv R_0$. It is assumed
that $\tau_{\nu_0}^\infty(J) = 0$.  This ignores the possible absorption by Sgr A
West of the low frequency ($\nu_0 < 10^9 $ Hz) radiation (Beckert et al.  1996). 
Sgr A West is an HII region surrounding Sgr A*.

\section{Parameters for a Spherical Accretion Model for Sgr A*}\label{sec:specparms}
The stellar winds that accrete onto Sgr A* are thought to be hypermagnetosonic, so
it is sensible to continue using the standard definition of an accretion radius
(Eq. [\ref{eq:radef}]), which is defined to be the point at which the gravitational
potential energy is equal to the {\sl initial} kinetic energy of the gas,
but using an average of the stellar wind velocities weighted by the mass loss rate.
For the Galactic Center, $R_A \sim 10^{16-17}$ cm.  However, this definition of
$R_A$ does not necessarily give the physical radius at which the capture always
occurs.  For example, wind-wind collisions, combined with radiative cooling, can
reduce the effective value of $v_w$.  For simplicity, we use this 
standard definition of $R_A$ to set the primary length scale
of the problem, and put $R_0 = f_RR_A$, where $f_R$ is a parameter of order unity.  

Now, $\dot M$ is likely to be a
fraction of the total GC wind, $\dot M_w$, which is estimated to be
$\sim 3\times10^{-3}\;M_\odot$ yr$^{-1}$ (Geballe et al. 1991).  However, 
more recent work suggests that the mass loss rate of at least some of the 
central massive stars is less than previously thought (Hanson et al. 1998), 
so that $\dot M_w$ may be smaller as well.  If the GC wind is 
dominated by a source some distance $D > R_A$ away from Sgr A*, then one 
would expect
\begin{equation}
\dot M = {{R_A^2 \dot M_w}\over{D^2}} = {{(2GM)^2 \dot M_w}\over{v_w^4 D^2}}\;.
\end{equation}
For example, if the $1000 \kms$ wind from IRS 13E1 (which is thought to be $\sim 0.16$ pc 
away from Sgr A*; see Melia \& Coker 1999),
dominates the flow, then $\dot M \sim 10^{21} \gms$.  Observationally, the mass accretion
rate is not very well constrained; a reasonable range is $\sim 10^{20-22}\gms$.
For all of the models presented here, we assume that the mass $M$ of the central black 
hole is $2.6\times10^6 \msun$, based on the latest observations (Eckart \& Genzel
1999; Ghez et al., 1998).  The magnetic field $B_0$ at 1 $R_A$ is thought to be a few milliGauss 
(Yusef-Zadeh et al. 1996) but its large scale average value may be a few times
smaller than this (Marshall et al. 1995).  In practice, we use a value for $B_0$ such that 
at $R_0$ the magnetic field is close to thermal equipartition:
\begin{equation}
B^2/8\pi = P_{mag} = \eta P_{thermal} = \eta 2nkT,
\end{equation}
where $\eta$ is a parameter of order unity.   Together, $\dot M$ and
$\eta$ effectively determine the normalization of
the model spectrum.

>From a combination of HeI line observations (Najarro et al. 1997) and hydrodynamic
modeling (Coker \& Melia 1997), the accreting wind is thought to be supersonic at $R_0$ with
a temperature $T_0 \sim 10^{6-7}$ K and a velocity $v_0 \sim 10^{2-3}\kms$.  
The temperature, $T_w$, and velocity, $v_w$, of the individual stellar winds that 
accrete onto Sgr A* are not necessarily the same as $T_0$ and $v_0$.  Wind-wind shocks 
and subsequent radiative cooling and acceleration result in $T_0 \simgt T_w$ and 
$v_0 \simlt v_w$.  The winds, which originate well outside of 1 $R_A$, are originally hypersonic
with a Mach number of 10-30, but by the time they reach 1 $R_A$, they are on average only
mildly supersonic with a Mach number of 1-3.  The non-zero velocity at infinity tends to make
$v_0 > v_{ff}$, where $v_{ff}$ is the free-fall velocity, while shocks tend to have the opposite
effect, making $v_0 \sim v_{ff}/4$.  Therefore, for simplicity, we assume that $v_0$ is
given by a fraction $f_v$, of order unity, of the free-fall value.
With $R_0 = f_RR_A$ and Equation (\ref{eq:radef}), we then have
\begin{equation}
v_0 = f_v v_w = f_v \sqrt{{{2 GM f_R}\over{R_0}}}\;.
\end{equation}
In order to end up with a supersonic accretion solution, it is also necessary to set
$T_0$ to be less than the local virial temperature, i.e., 
\begin{equation}
{{2 k T_0}\over{m_p}} < {{G M}\over{2 R_0}} \equiv {{2 k T_{vir}}\over{m_p}}\;.
\end{equation}
We parameterize the starting temperature according to the prescription 
$T_0 = f_T T_{vir}$, where again $f_T$ is of order unity.

\subsection{Angular Momentum Considerations}

Some accretion models of Sgr A* (e.g., Narayan et al. 1995) assume that 
$v_0 \ll v_{ff}$, requiring a deceleration mechanism fairly close
to 1 $R_A$.  Since $T_0$ and $B_0$ are too small to decelerate the inflowing
gas, the most likely reason for $v_0$ being so small would probably be 
that the gas contains a greater specific angular momentum than is seen in
the hydrodynamic simulations (Ruffert \& Melia 1994; Coker \& Melia 1997).
With the wind originating from more than a dozen sources distributed
fairly isotropically around Sgr A* (Melia \& Coker 1999), the time averaged 
specific angular momentum accreted by the black hole is likely to be small.  
This is the conclusion to which one is led with the current series of hydrodynamic
simulations.  The circularization radius at which the accreted specific 
angular momentum is equal to its Keplerian value is
\begin{equation}
R_{circ} =2\lambda^2r_s\;,
\end{equation}
where $r_s\equiv 2GM/c^2$ is the Schwarzschild radius
and $\lambda$ is the accreted specific angular momentum in units of $c r_s$.
The hydrodynamic simulations suggest that $\lambda$ is highly time variable
with a value in the range $2-50$.  However, these calculations have had limited
physical resolution near the black hole and they have therefore sampled the
$\lambda$ accreted within the region $R_{boundary} \gg r_s$.  As one would expect,
tests have shown that smaller values of $R_{boundary}$ result in smaller 
values of $\lambda$ carried across that threshold.  Thus, the average
values for $\lambda$ given above may be thought of as upper limits so that $R_{circ} \ll R_A$.
Similarly, the viscous transport of angular momentum outward and a possible
mass loss via a wind off the disk (see, e.g.,
Xu \& Chen 1997) should result in $\dot M(R_{circ}) > \dot M(r_s)$.  Numerical 
calculations as yet do not have the physical resolution to verify this in the 
case of Sgr A*.

More importantly, these simulations indicate that the vector {\sl direction} of the accreted
specific angular momentum is highly time variable and dependent on conditions outside of the
accretion radius, with sign changes occurring on a time scale of 1-50 years.  Thus,
with $R_{circ}$ representing the length scale within the nascent disk (though
the distribution of angular momentum outwards may increase the disk's outer radius
somewhat above this value), even if
such a disk forms, it is not likely to be stable over a time scale of more than a few decades,
since the gas contains no long term preferred angular momentum axis.
Recent work (Genzel 1998) suggests that the wind sources may have a small 
net rotation around Sgr A* and, if the resulting additional angular momentum is not 
canceled in wind-wind collisions, then angular momentum could play a more significant 
role in the spectrum of Sgr A* than we assume here.

\subsection{The Magnetic Field}

In this paper, we do not attempt (as Kowalenko \& Melia 1999 did) to determine
the magnetic dissipation rate from first principles, but rather we will use 
our fits to the observed spectrum of Sgr A* to infer empirically what the
profile of $B$ should be in order to validate this picture.  In other words,
we have a sense of how $B$ might behave (most likely different from simple
equipartition) based on the calculations of Kowalenko \& Melia (1999), and
we use that algorithm for $B(r)$ to fit the spectrum.  We find that a 3 component profile
is necessary and sufficient for this purpose.  The profile of $B$ affects primarily 
the heating term, $\Gamma$, and the magnetic bremsstrahlung emissivity, $j_\nu^{sync}$,
as discussed above.  With 3 components, we have 5 parameters describing the magnetic
field:  3 power-law indices ($p_1, p_2, p_3$) and the location of 
2 breakpoints ($r_1, r_2$).  Since the magnetic field of the
ISM is observed to be approximately in thermal equipartition, we fix the first index, $p_1$,
so that  $B(r) \propto r^{-5/4}$, or $P_{mag} = \eta P_{ram}$.  The shape of the final
spectrum is sensitive to the remaining 4 parameters ($r_1$, $p_2$, $r_2$, and $p_3$) but
we find that one general configuration works best.  The first
breakpoint $r_1$ occurs at a few hundred $r_s$, with $B(r)\rightarrow$ constant (i.e.,
$p_2 = 0$).  Interestingly, the second breakpoint $r_2$ is at $\sim 3 r_s$, the
radius of the last stable orbit for a non-rotating black hole.  At that point, $B(r)
\propto r^{-3-{-6}}$, which suggests the presence of a magnetic dynamo at small radii.

These three segments of $B(r)$ (an equipartition region, a region of constant field
intensity, and a dynamo region) are all necessary to fit the general flat spectral shape 
of Sgr A*.  The equipartition region results in a sufficiently large field and 
temperature to reproduce the observed flux at low frequencies (i.e.,
$10^9 \simlt \nu \simlt 10^{10}$ Hz) with $r$ large enough so that the
emission is not self-absorbed.  The region of constant magnetic field is required to match
the flat spectral index of $\sim 0.3$ (Lo 1986) at moderate radio frequencies
($10^{10} \simlt \nu \simlt 10^{11}$ Hz).  At $\sim2\times10^{11}$ Hz there is
an observed sub-millimeter excess, presumably due to an ultra-compact component a few $r_s$
in size (Falcke et al. 1998); our dynamo region produces this excess.  There are many 
arguments (see, e.g., Hawley \& Balbus 1992) for the presence of a magnetic dynamo 
close to a black hole so such a component is not implausible.  

The calculation of the magnetic bremsstrahlung contribution to
$j_{\nu}$ assumes that $\mu$, the cosine of the angle between the flow and the line 
of sight, is the same as the cosine of the angle between $B$ and the line of sight.
For a spherical flow, in the absence of magnetic reconnection (which should be minimal 
for $r \gg R_A$ anyway), $B_r \propto r^{-2}$, while $B_\theta \propto
r^{-1}$. Thus, even if the magnetic field at infinity is perfectly tangled, it is 
reasonable to assume that at $R_0$, $B_r \gg B_\theta$.  However, because the 
magnetic field is divergence-less, the stretched out field lines must 
close somewhere.  Presumably this occurs quite close to the black hole. In a fully
self-consistent calculation, we will use the distributions derived from an
actual 3D hydrodynamic simulation as the basis for calculating the emissivity.

\section{Results}\label{sec:specres}

\begin{figure}[p]
  \centering
  \psfig{figure=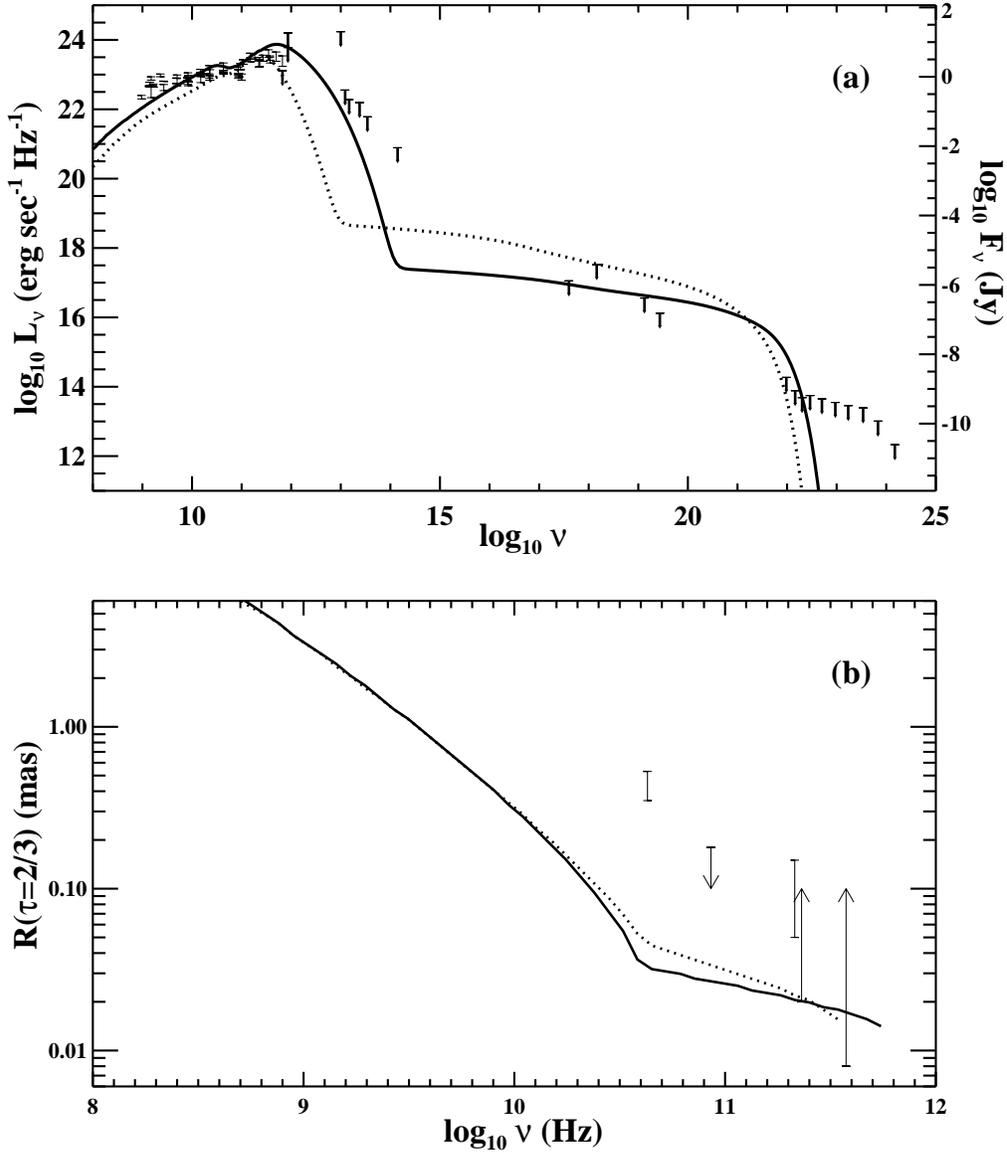,width=5.2in}
  \caption[Comparison of predicted Sgr A* spectrum and size with observations.]
    {The solid curves are for a model which assumes pure free-fall ($H\equiv 0$) while the dotted curves
    are for a model which uses the complete equations (see text for details).
    (a) The observed spectrum of Sgr A* along with the predicted results for the best model fits.
    The observational data are taken from various sources (see Narayan et. 
    al. 1998 for a recent compilation).
    (b) The observed size of Sgr A* along with the predicted size, 
    defined as the radius at which the optical depth, $\tau_{\nu_0}^\infty(r)$ equals 2/3.}
  \label{fig:specsize}
\end{figure}

We present the resulting spectrum of our best fit model in Figure \ref{fig:specsize}(a).
For comparison with previous calculations (e.g., Melia 1994), we also plot results for a
model which assumes only gravity acts to accelerate the accreting
gas (i.e., $H\equiv 0$).  Since the flow is only mildly supersonic at 
moderate radii, the assumption of free-fall is physically invalid, but such 
models do allow hotter gas to exist at larger radii and, thus, have more 
emission at lower frequencies than the full solution models.
Nonetheless, both model spectra are consistent with the observations
over 16 decades of frequency.  The parameters used in the two fits are given in
Table \ref{tab:specparms}.  Note that one clear prediction of these models is
that the source of the majority of the high energy $\gamma$-ray emission
is not Sgr A* or its associated accretion flow, unless some
exotic particle acceleration mechanism is operating within the inflow.  
This is consistent with recent models (Melia, Yusef-Zadeh, \& Fatuzzo 1998),
which suggest the GC high energy emission is due to Sgr A East, a 
super-nova-like remnant near or possibly even enveloping Sgr A*.  The details of the predicted emission
in the IR region are fairly sensitive to the details of radiative transport, 
since the gas circularizes at small radii, possibly producing a small
disk which emits primarily in the IR.  Nonetheless,
the next generation of infrared observations should be 
able to detect the emission from the inner $\sim 10 r_s$ of the black hole.

\begin{table}
  \centering
  \caption{Parameters for the Sgr A* Spectral Fits}
  \label{tab:specparms}
  \vspace{0.3in}
  \begin{tabular}{|l|c|c|}

\hline
                         &               &               \\ [-11pt]
Parameter                & Full Solution & Free-fall     \\
                         &               &               \\ [-11pt]
\hline
                         &               &               \\ [-11pt]
$\mdot ({\rm g sec^{-1}})$&8\E{21}       & 5\E{21}       \\
$v_\infty (\kms)$        & 500           & 1000          \\
$f_R$                    & 1.0           & 1.0           \\
$f_T$                    & 0.5           & 1.0           \\
$f_v$                    & 0.5           & 1.0           \\
$\eta$                   & 0.05          & 0.15          \\
$p_1$                    & 2.5           & 2.5           \\
$p_2$                    & 0.0           & 0.0           \\
$p_3$                    & 6.0           &12.0           \\
$r_1 (r_s)$              & 750           & 2000          \\
$r_2 (r_s)$              & 4             & 3             \\
 \hline
  \end{tabular}
  \parbox{5.25in}{
  \vspace{0.25in}
  Model parameters for two fits to the spectrum of Sgr A*, assuming spherical 
  accretion.  The full solution is for a calculation using the equations described 
  in the text while the free-fall model assumes $H\equiv0$ so that only gravity 
  accelerates the infalling gas (see Eq. [\ref{eq:vprime}]).
  }
\end{table}

Some of the observations used in Figure \ref{fig:specsize}(a) are averaged over
a span of years while others are contemporaneous.
Since the flux of Sgr A* can vary with time by more than $25\%$ (depending on
frequency), we do not necessarily wish to reproduce the entire
suite of observations but rather the general trend.
The resolution of the observations also varies greatly but most of the values reported in the
table have backgrounds subtracted out and in principle refer to the flux
of Sgr A* only.  Those observations which have a potential for source confusion are plotted as upper
limits.  For example, 
at $\nu \simgt 8.6\times10^{11}$ Hz, the background emission due to, e.g., dust at a temperature
of 90K, is
comparable to emission from Sgr A* itself (Zylka et al. 1992).  This is particularly true in the
infrared, where Sgr A* has yet to be definitely identified and has a flux considerably
less than that of the surrounding gas and stars, and in the X-rays and $\gamma$-rays,
where the poor spatial resolution is likely to result in multiple sources being in the field of view.
Similarly, observations at $\nu \simlt 1$ GHz are unreliable as limits for the luminosity
of Sgr A* due to source
confusion, scattering, and self-absorption (Davies et al. 1976).
In converting the observed fluxes to the luminosities plotted in
Figure \ref{fig:specsize}(a), it is assumed that Sgr A* is 8.5 kpc away.

The predicted size of Sgr A* is shown in Figure \ref{fig:specsize}(b).  Also shown are
the present observational limits and measurements (see Lo et al. 1998 for a summary
of recent work).  At present there are two observations
which address the intrinsic diameter of the minor axis of Sgr A*.
However, these observations are difficult and, for example, may have had
difficulties with calibration (see, e.g., Krichbaum et al. 1998).  The model
satisfies the upper limit at 3.5mm as well as the lower limit at 0.8mm,
but it predicts a source that is somewhat more compact than the other observations
suggest.  But while the observations tend to fit Gaussian FWHM to the data, the predicted 
diameter is defined as the last scattering surface (i.e., $r[\tau=2/3]$).  For a more 
detailed and accurate size comparison, one needs a more sophisticated treatment,
using, for example, the theory of wave propagation in an extended, irregular medium (Melia,
Jokipii \& Narayanan 1992).  In addition, at small radii, the presence of angular momentum
and asymmetries in the flow will result in a distended non-circular source shape.  
For example, at 7mm the axial ratio is observed to be less than $0.3$ (Lo et al. 1998).  
Thus, the model size results presented here should be taken as approximate lower limits.

\begin{figure}[p]
  \centering
  \psfig{figure=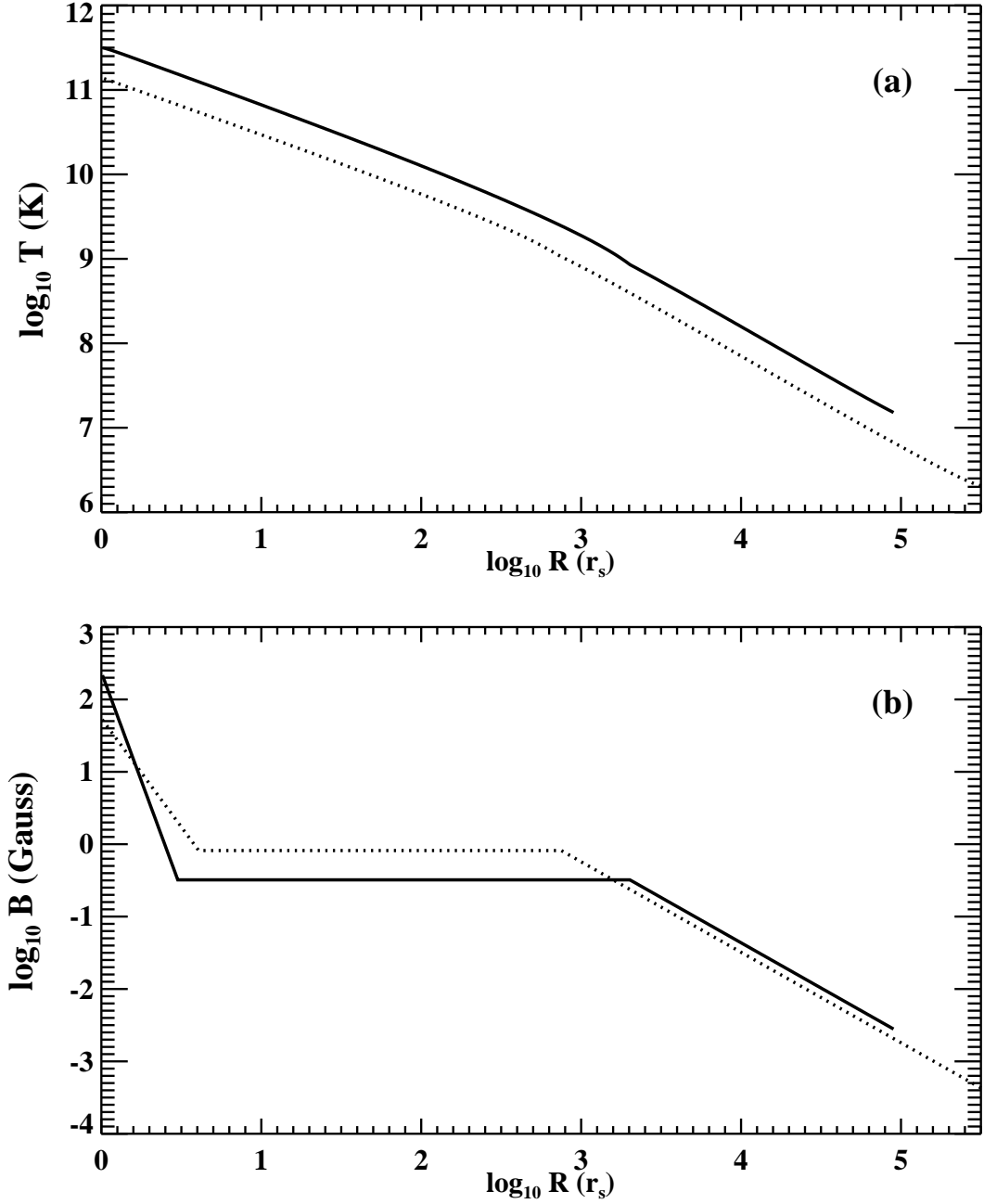,width=5.85in}
  \caption[Plot of temperature and magnetic field profiles for the best fit models.]
    {Plots of temperature (a) and magnetic field (b) versus radius for the two best fit models.
    The solid curves are for a model which assumes pure free-fall ($H\equiv 0$) while the dotted curves
    are for a model which uses the complete equations (see text for details).}
  \label{fig:specTandB}
\end{figure}

In Figure \ref{fig:specTandB} we show the radial profiles for the temperature 
and magnetic field for the two fits shown in Figure \ref{fig:specsize}.  The gas 
is always sub-virial ($T_{vir}\sim10^{12}\K\;\,[r/r_s]^{-1}$)
with a peak temperature of $\sim 10^{11}$K.  This peak temperature is within
the present observed lower ($1.3\times10^{10}$K; Lo et al. 1998) and upper
($5\times10^{11}$K; Gwinn et al. 1991) brightness temperature limits.
Only within $\sim 10^{2-3} r_s$
of the black hole, where the gas becomes relativistic, is
there significant magnetic bremsstrahlung
emission.  It is not a coincidence that $r_1$ is $\sim 10^3 r_s$.
The accreting gas must stay hot enough at such radii to produce the flat spectrum; 
the constant magnetic field and subsequent heating via reconnection very nearly achieve this.  
However, this cannot continue all the way down to the event horizon or else the 
resulting magnetic field is insufficient to produce the observed sub-millimeter 
excess and the gas gets too hot to continue accreting and becomes pressure
supported.  Thus, we require a sharp increase in the magnetic field strength 
within $\sim 10 r_s$ of the black hole with a resulting peak magnetic field 
of $\sim 100$ Gauss.  We postulate that this may occur when the frozen-in 
magnetic field lines, which have been stretched out during the accretion 
process, finally close or are twisted by sheared gas motions due
to a transition resulting from residual angular momentum in the flow; however, if
the viscosity is large, the bulk gas flow still remains primarily radial and supersonic (Narayan et al. 1997).
As described by Hawley and Balbus (1992), instabilities in the flow, possibly tied to differential
rotation once the gas circularizes, may result in the generation of
strong poloidal and toroidal fields, driving the magnetic field strength up to thermal
equipartition, as required in our model.
Except in the transrelativistic region ($\sim 10^3 r_s$), the gas density and velocity
profiles are close to that from free-fall so that n$\sim 10^{10} {(r_s/r)}^{3/2}\ncm3$.

\section{Conclusions}\label{sec:specdis}

By incorporating an enhanced treatment of the magnetic bremsstrahlung emissivity and
solving the accretion flow equations explicitly, we have improved the accreting black
hole model for Sgr A* with resulting spectra that are consistent with observations over
more than 16 decades of frequency.  The mass accretion rate of $\sim 10^{-4} \msun $yr$^{-1}$,
as determined
from the spherical accretion model, is
consistent with the rate expected on the basis of observations and hydrodynamical
arguments (see, e.g, Coker \& Melia 1997).  However, other models, such as an
``advection dominated accretion flow'' (ADAF) with outflow (Blandford \& Begelman 1999),
require an accretion rate that is 2 or more orders of magnitude smaller than this.

There are some difficulties with the model.  The full solution, using the Euler
equation, suggests that the flow may be transonic and that it may therefore
shock at smaller radii. A shock could produce high energy particles, 
which would result in significant magnetic bremsstrahlung emission at larger 
radii (Markoff, Melia, \& Sarcevic 1997), thereby increasing the emission
at lower frequencies.  We hope to address this difficulty in future work.  
It is interesting that the break seen in the low energy $\gamma$-ray observations
coincides with the high energy turnover in the model, suggesting the presence
of another, hotter source within the field of view of EGRET.
In addition, we have effectively ignored the role played by angular momentum, 
the presence of which would introduce viscous processes that are even more 
poorly understood than magnetic reconnection.  Also, the handling of
radiative transport has so far been very simplistic.  Inclusion of
relativistic 3D radiative transfer, including ray bending and inverse
Compton, could potentially alter the resulting spectra significantly.

The model suggests that if the accretion rate onto Sgr A* is large
($\sim 10^{-4} \msun $yr$^{-1}$) then the observed X-ray emission from
the GC is due to the extended X-ray emission from this accretion process.
However, the observations have a resolution on the order of arcminutes
while the `extended' emission region from the model spans a few arcseconds.
Due to calibration difficulties and an uncertain column depth between here
and the GC, the observed X-ray limits are uncertain to within a factor of a
few (Narayan et al. 1998) so the model is still marginally consistent with
the X-ray observations.  The next generation of observations (such as with Chandra)
will most likely settle this issue.

The picture with which we have worked here under-predicts somewhat the low frequency 
radio emission.  Within the
context of the model, unreasonable temperature and magnetic field profiles
are required to fit the observations and even then the fit at other
frequencies degrades.  This difficulty is due to two things.  First, the low
frequency emission from close to the black hole is trapped so it does not contribute
to the low frequency spectrum.  Second, gas with sufficient magnetization 
and temperature to produce the observed magnetic bremsstrahlung emission
at large radii is gravitationally unbound and could not self-consistently accrete.
An improvement to the model that addresses this low-$\nu$ deficiency in the
spectrum may come with more realistic 3D simulations, which will be
reported elsewhere.

Finally, although we have briefly touched on the likely importance of the 
sub-millimeter excess to our understanding of the environment just outside 
the event horizon, this clearly is an issue that warrants further detailed
theoretical work.  Taking into account scatter-broadening of the 
image in the interstellar medium, and the finite achievable telescope
resolution, the $\sim10 r_s$ ``shadow'' of the Galactic Center black hole
(see, e.g., Falcke, Melia \& Agol 1999) should be well 
observable with very long-baseline interferometry at sub-millimeter wavelengths. 
In our picture, the dynamo effect that leads to the intensification of the
magnetic field that accounts for this spectral excess 
may be due to the circularization of the infalling gas when it approaches 
the circularization radius $2\lambda^2r_s$ (see \S\ 3.1).  We are in the process
of examining the behavior of the magnetized plasma when shearing motions become
important in this region, and we will report on the results of
this investigation in the near future.

\end{document}